\documentclass[11pt,a4paper]{article}

\usepackage[utf8]{inputenc}
\usepackage[T1]{fontenc}
\usepackage{lmodern}
\usepackage[margin=25mm]{geometry}
\usepackage{microtype}

\usepackage{amsmath,amssymb}
\usepackage{graphicx}
\usepackage[font=small,labelfont=normalfont]{caption}

\usepackage{authblk}
\usepackage[numbers,square,sort&compress]{natbib}
\usepackage[hidelinks]{hyperref}
\hypersetup{
    pdftitle={Robust Active Learning for More Efficient and Reliable Paths to Optimal Design},
    pdfauthor={Hayriye Pehlivan Solak, Francesco Di Fiore, Jeroen Wackers, Riccardo Pellegrini, Andrea Serani, Matteo Diez, Laura Mainini}
}

\newcommand{\manuscriptfigure}[2][]{%
    \IfFileExists{#2}{%
        \includegraphics[#1]{#2}%
    }{%
        \PackageWarningNoLine{main_arxiv}{Missing figure #2; supply the original before submission}%
        \fbox{\parbox[c][35mm][c]{0.86\linewidth}{%
            \centering\small Figure file not supplied.\\[0.5em]
            \texttt{\detokenize{#2}}%
        }}%
    }%
}

\newcommand{\DesignVector}{\ensuremath{\mathbf{x}}}

\newcommand{\CandidateSet}{\ensuremath{\mathcal{X}_{\mathrm{cand}}}}
\newcommand{\PositivePart}[1]{\ensuremath{\left[#1\right]_{+}}}

\newcommand{\LF}{\ensuremath{\mathrm{L}}}
\newcommand{\HF}{\ensuremath{\mathrm{H}}}
\newcommand{\LFHF}{\ensuremath{\mathrm{L{+}H}}}

\newcommand{\ObservedSitesLF}{\ensuremath{\mathcal{X}_{\LF}^{\mathrm{obs}}}}
\newcommand{\ObservedSitesHF}{\ensuremath{\mathcal{X}_{\HF}^{\mathrm{obs}}}}

\newcommand{\MeanHF}[1]
{\ensuremath{\mu_{\HF}\!\left(#1\right)}}

\newcommand{\InterpolationUncertaintyLF}[1]
{\ensuremath{U_{\mathrm{int},\LF}\!\left(#1\right)}}

\newcommand{\UncertaintyHF}[1]
{\ensuremath{U_{\HF}\!\left(#1\right)}}

\newcommand{\BestObservedHF}
{\ensuremath{f_{\HF}^{\min}}}

\newcommand{\PredictedImprovementGap}[1]
{\ensuremath{d\!\left(#1\right)}}

\newcommand{\TotalPotential}[1]
{\ensuremath{P\!\left(#1\right)}}

\newcommand{\CostLF}{\ensuremath{c_{\LF}}}
\newcommand{\CostHF}{\ensuremath{c_{\HF}}}
\newcommand{\CostLFHF}{\ensuremath{c_{\LFHF}}}

\title{Robust Active Learning for More Efficient and Reliable Paths to Optimal Design}

\author[1,2,3]{Hayriye Pehlivan Solak\thanks{Corresponding author: \href{mailto:h.pehlivansolak@imperial.ac.uk}{h.pehlivansolak@imperial.ac.uk}.}}
\author[1,2]{Francesco Di Fiore}
\author[3]{Jeroen Wackers}
\author[4]{Riccardo Pellegrini}
\author[4]{Andrea Serani}
\author[4]{Matteo Diez}
\author[1,2]{Laura Mainini}

\affil[1]{Brahmal Institute for Sustainable Aviation, Imperial College London, United Kingdom.}
\affil[2]{Department of Aeronautics, Imperial College London, United Kingdom.}
\affil[3]{LHEEA Lab, Centrale Nantes / CNRS UMR 6598, Nantes, France.}
\affil[4]{National Research Council-Institute of Marine Engineering, Rome, Italy.}

\date{}

\begin{document}
\maketitle

\begin{abstract}
Simulation-driven design optimization requires active learning strategies that efficiently combine models with different computational costs and levels of fidelity. We introduce Multifidelity Cost-Aware Screening (MF-CAS), which enables the joint selection of sampling location and fidelity using a surrogate that provides decomposed uncertainty intervals rather than a complete predictive probability distribution. MF-CAS exploits the contribution of each fidelity to the high-fidelity uncertainty together with its computational cost to determine whether a candidate should first be evaluated at low fidelity or directly progressed towards high fidelity. The method is assessed on the constrained 14-dimensional hydrodynamic optimization of the DTMB 5415 hull under a deliberately limited computational budget. MF-CAS achieves an approximately $9.0\%$ resistance reduction, compared with $6.7\%$ for the best baseline optimisation method, demonstrating that incorporating fidelity-resolved uncertainty directly into the joint location-fidelity decision can substantially improve multifidelity active learning efficiency.

\end{abstract}


\section{Introduction}
Simulation-driven design optimization plays a central role in the development of high-performing engineering systems, but its practical application is often limited by the computational cost of high-fidelity numerical models \cite{serani2021urans, serani2024scoping}. Multifidelity methods reduce this burden by combining inexpensive, low-fidelity evaluations, which support broader exploration of the design space, with a limited number of high-fidelity evaluations to refine the design solution \cite{mainini2024lecture}. Their efficiency, however, strongly depends on the active learning strategy adopted to determine where the next evaluation should be performed and at which fidelity \cite{mainini2026arco}. Many active learning strategies rely on surrogate-model uncertainty to balance the exploration of poorly known regions against the exploitation of designs predicted to improve the current solution \cite{volpi2015development}.

The reliability of the sample selection is particularly important when the available data are limited or subject to numerical noise, as is common in simulation-driven engineering applications \cite{pellegrini2023multifidelity}. Our recent work on multifidelity surrogate modelling addressed this issue through an uncertainty estimation that separates interpolation and training-point uncertainty and propagates their contributions through the multifidelity surrogate \cite{wackers2024adaptivity}. This formulation provides rich information on both the magnitude and origin of the uncertainty affecting the high-fidelity prediction. It only predicts uncertainty intervals however, which prevents the direct application of acquisition functions based on predictions of full probability distributions, such as expected improvement or probability of improvement \cite{karen2018siam}. Lower confidence bound strategies remain applicable because they operate directly on the surrogate prediction and uncertainty. In their multifidelity formulation, however, the sampling location is first selected from the high-fidelity prediction and total uncertainty, while the fidelity is determined only afterward from the individual uncertainty contributions and their computational costs \cite{pellegrini2022derivativefree}. The location is therefore fixed before the benefit of sampling at a specific fidelity is considered, potentially overlooking a different location where another fidelity provides greater value for the available computational expenditure.

To address this limitation, this work introduces MF-CAS (Multifidelity Cost-Aware Screening), an uncertainty-driven multifidelity active learning strategy in which the fidelity-dependent uncertainty contributions are incorporated directly into the selection of the sampling location. For each candidate design, the proposed formulation evaluates whether it should be sampled at low fidelity, high fidelity, or both fidelities by considering (i) how strongly each fidelity contributes to the uncertainty of the high-fidelity prediction, (ii) the potential improvement of the objective indicated by the high-fidelity surrogate, and (iii) the corresponding computational cost. Location and fidelity are therefore evaluated simultaneously, allowing the method to select the combination that provides the greatest value for the high-fidelity search. The method is assessed on the constrained hydrodynamic shape optimization of the DTMB 5415 hull using a potential-flow model, where low- and high-fidelity levels use a coarse or fine panel distribution on the hull surface.

\section{Multifidelity active learning framework}\label{sec:MF}
The proposed multifidelity active learning framework combines two main elements: an adaptive multifidelity surrogate model, which synthesizes observations obtained from models with different levels of fidelity and provides the corresponding prediction and uncertainty estimation; and an active learning policy, which uses this information to select the next design and fidelity action during the optimization process. Starting from an initial set of evaluations, the surrogate is iteratively updated as new observations are acquired, while the sampling policy guides the allocation of the available computational budget towards the identification of improved high-fidelity designs. The multifidelity surrogate and its uncertainty formulation are introduced first, followed by the proposed active learning strategy.

\subsection{Multifidelity Surrogate and Uncertainty Estimation} \label{sect:MFUE}
\paragraph{Surrogate model} The surrogate model used in the multifidelity setting is based on Stochastic Radial Basis Functions \cite{pellegrini2023multifidelity}; given $\mathbf{x}\in\mathcal{X}\subset\mathbb{R}^D$ as a design variable vector of dimension $D$, an objective function $f(\mathbf{x})$ is assessed by two fidelity levels: the high-fidelity model $f_H(\mathbf{x})$ and the low-fidelity model $f_L(\mathbf{x})$. Observations {$s_l(\mathbf{x})$} (with $l\in [L,H]$) can be perturbed by (fidelity-dependent) Gaussian white noise: $s_l(\mathbf{x}) = f_l(\mathbf{x}) + \mathcal{N}(0,\sigma_{n_l})$. The multifidelity (MF) prediction $\widehat{f}(\mathbf{x})$ of $f(\mathbf{x})$ is the sum of the surrogates (denoted by $\, \widetilde{\cdot} \,$) for the low fidelity and the inter-level error \cite{pellegrini2023multifidelity}:
$
    \label{eq:NFPred}
    \widehat{f}(\mathbf{x})=\widetilde{f}_L(\mathbf{x})+\widetilde{\varepsilon}(\mathbf{x}).
$
The two training sets are $\mathcal{T}_l=\{{\bf {x}}_{i},s_l({\bf {x}}_{i})\}_{i=1}^{J_l}$, with $J_l$ the training set size. Nested data are considered: 
$\{{\bf {x}}_{i}\}_{i=1}^{J_H} \subseteq \{{\bf {x}}_{i}\}_{i=1}^{J_{L}}$.
The inter-level error training set is defined as $\mathcal{E}=\{{\bf {x}}_{i}, \varepsilon(\mathbf{{x}}_{i}) \}_{i=1}^{J_H}$, where 
$\label{eq:errors}
    \varepsilon(\mathbf{{x}}_{i})=s_H(\mathbf{{x}}_{i})-\widetilde{f}_{L}(\mathbf{{x}}_{i}). 
$
Each prediction $\widetilde{\cdot}_l\left({\bf x}\right)$ is computed as the expected value (EV) over a stochastic tuning parameter $\tau\sim \textrm{unif}[1,3]$ of a surrogate model $\widetilde{g}_l$ following the principle in \cite{volpi2015development}:  
\begin{equation}\label{eq:LSSRBF} 
    \widetilde{f}_l\left({\bf x}\right)=\mathrm{EV}\left[\widetilde{g}_l\left({\bf x},\tau\right)\right]_\tau, \quad \text{with} \quad
    \widetilde{g}_l\left({\bf x},\tau\right)=\mathrm{EV}\left[\mathbf{s}_l\right]+\sum^{{M_l}}_{j=1} w^l_j ||{\bf x}-{\bf c}^l_j||^{\tau},
\end{equation}
where $w^l_j$ are unknown coefficients, $||\cdot||$ is the Euclidean norm, {$\mathbf{s}_l=\{s_l(\mathbf{x}_i)\}_{i=1}^{J_l}$} and $\mathbf{c}^l_j$ are the RBF centres, with $j=1,\dots,M_l$ and ${M}_l\leq J_l$.  For each fidelity level, if the data are not affected by numerical noise ($\sigma_{n_l}=0$), exact interpolation of the training set is imposed and the weights $w_j^l$ are computed by solving $\mathbf{A} {\bf w}^l=(\mathbf{s}_l-\mathrm{EV}\left[\mathbf{s}_l\right])$, with ${\bf c}_j^l={\bf x}_j$ (i.e.\ $M_l=J_l$). In the presence of noise, a number of RBF centres ${M}_l$ is chosen smaller than the number of training points ${J}_l$, and ${\bf c}_j^l$ coordinates are defined via $k$-means clustering of the training point coordinates. The number of RBF kernels $M_l$ ranges from $2D$ to $J-1$; least-squares regression provides more smoothing of the noisy training set when the number of centres $M_l$ is small. The optimal number of stochastic RBF centres is defined by minimizing a leave-one-out cross-validation (LOOCV) metric. Finally, the weights $w_j^l$ are determined with least-squares regression (LS) by solving $\mathbf{A}^{\sf T} \mathbf{A}{\bf w}^l= \mathbf{A}^{\sf T} (\mathbf{s}_l-\mathrm{EV}\left[\mathbf{s}_l\right])$.

\paragraph{Uncertainty estimation} The uncertainty is modeled \cite{wackers2024adaptivity} by dissecting it into three main components: i) interpolation uncertainty under sparse data, ii) noise-filtering, and iii) multifidelity propagation. 

The SRBF uncertainty $U^{\tilde{g}}(\mathbf{x})$ is the 95\% confidence interval over the predictor family $g_l(\mathbf{x},\tau)$ of Eq.~\eqref{eq:LSSRBF}. This estimation is only accurate if all fits capture the true function reasonably well. But in the early stages of design optimisation, the data are too sparse to support a statistically consistent estimate, so a baseline uncertainty is imposed based on the assumption that any function consists of peaks and valleys with a characteristic width $2r_0$ and height $P_{U_0}$. Considering the domain is normalised, the peak width is taken as $0.25$, and the peak height as the data variation $P_{U^0} = \max_i s(\mathbf{x}_i) - \min_i s(\mathbf{x}_i)$. A parabolic behaviour assumption depending on the closest data point is used for defining the baseline uncertainty, which is switched off once the surrogate starts to have enough data to capture the second derivatives (at least $2D+1$ points in the sphere $r_0$ around $x$). For counting, the sphere radius is defined with a smeared Heaviside function, for a smooth transition between points inside and outside the $r_0$ sphere.

Noise-filtering introduces two uncertainty sources. First, the noise magnitude is unknown; therefore, noise-filtered data are obtained as a mean of least-squares fits using different kernel counts $M_l$, with likelihood of each $M_l$ based on LOOCV errors. Second, the local mean of the data may deviate from the actual function. We estimate this uncertainty with the central limit theorem applied to the local mean, requiring an estimate of the noise level and the count of points contributing locally to the filtered function (via $k$-means clustering of points to RBF centres). To avoid underestimation, the noise level is taken as the largest value for which the filtered result remains within the 95\% interval.

For multifidelity, uncertainty propagates from the low to the high fidelity conditionally on the HF data, since the uncertainty contribution from the LF surrogate in the HF training points is zero. Conditional training-point uncertainties follow the two-point correlation of noise-filtered data as a function of the local distances between RBF kernels, where the correlation function is chosen as a parabola similar to the baseline uncertainty.

\subsection{MF-CAS: Multifidelity Cost-Aware Screening}

The uncertainty formulation introduced above provides an uncertainty interval for the high-fidelity prediction together with the contributions associated with the different components of the multifidelity surrogate. Since it does not provide a complete predictive probability distribution, acquisition functions that explicitly require such a distribution cannot be applied directly \cite{takeno2020multifidelity,ruan2020variablefidelity}. LCB-based strategies remain applicable, but multifidelity LCB formulations identify the sampling location from the high-fidelity prediction and total uncertainty, while the fidelity is subsequently selected from the individual uncertainty contributions relative to their computational costs \cite{pellegrini2022derivativefree}. This separation can strongly favour low-fidelity evaluations when their uncertainty-to-cost ratio is large, even when acquiring such information has limited relevance to the high-fidelity optimization. We introduce Multifidelity Cost-Aware Screening (MF-CAS) to remove low-fidelity sampling from this direct competition and instead evaluate each candidate design according to the high-fidelity opportunity that can be accessed and the cost required to obtain it.

For the optimization setting considered here, let \BestObservedHF{} denote the best available high-fidelity objective value. For each candidate design \DesignVector, the improvement gap and the corresponding high-fidelity optimization potential are defined as
\begin{equation}
\PredictedImprovementGap{\DesignVector}=\BestObservedHF-\MeanHF{\DesignVector}, \qquad
\TotalPotential{\DesignVector}=\PositivePart{\PredictedImprovementGap{\DesignVector}+\UncertaintyHF{\DesignVector}},
\label{eq:HF_potential}
\end{equation}
where $\left[z\right]_{+}=\max(z,0)$. The quantity \TotalPotential{\DesignVector} is therefore positive when improvement is either predicted by the surrogate or remains compatible with its uncertainty interval, and provides the common measure used to value access to the high-fidelity response.

For two fidelity levels, the cost of obtaining this response depends on the information already available. If the candidate design has previously been evaluated with the low-fidelity model, only a high-fidelity evaluation is required. At a new candidate design, nested data require both low- and high-fidelity evaluations. MF-CAS consequently defines
\begin{equation}
A_{\HF}(\DesignVector)=\frac{\TotalPotential{\DesignVector}}{\CostHF}, \qquad
A_{\LFHF}(\DesignVector)=\frac{\TotalPotential{\DesignVector}}{\CostLFHF}, \qquad
\CostLFHF=\CostLF+\CostHF.
\label{eq:mfcas_hf_scores}
\end{equation}
The first score applies to $\DesignVector\in\ObservedSitesLF\setminus\ObservedSitesHF$, whereas the second applies to $\DesignVector\in\CandidateSet\setminus\ObservedSitesLF$. Importantly, $A_{\LFHF}$ does not assign the cost $\CostLFHF$ to a low-fidelity simulation. It measures the opportunity per unit cost of obtaining a high-fidelity response at a previously unsampled design, for which both fidelity levels are required by the nested surrogate.

At a new candidate design, MF-CAS then determines whether the low-fidelity evaluation should be performed separately before the high-fidelity cost is incurred. The low-fidelity interpolation uncertainty \InterpolationUncertaintyLF{\DesignVector} is the component directly affected by observing the low-fidelity response at that design. Removing this contribution in quadrature gives the residual high-fidelity uncertainty, potential, and opportunity
\begin{equation}
\begin{aligned}
U_{\mathrm{H}\setminus\mathrm{L}}(\DesignVector)
&=\sqrt{\max\left(\UncertaintyHF{\DesignVector}^{2}-\InterpolationUncertaintyLF{\DesignVector}^{2},0\right)}, \\
P_{\mathrm{H}\setminus\mathrm{L}}(\DesignVector)
&=\PositivePart{\PredictedImprovementGap{\DesignVector}+U_{\mathrm{H}\setminus\mathrm{L}}(\DesignVector)}, \\
A_{\mathrm{H}\setminus\mathrm{L}}(\DesignVector)
&=\frac{P_{\mathrm{H}\setminus\mathrm{L}}(\DesignVector)}{\CostHF}.
\end{aligned}
\label{eq:mfcas_residual}
\end{equation}
This construction is not intended to predict the posterior uncertainty after the low-fidelity observation. Rather, it estimates how competitive the candidate design would remain if the interpolation uncertainty that can be examined through low-fidelity evaluation were excluded from its current high-fidelity opportunity.

The residual opportunity is compared with the strongest competing opportunity that would remain affordable after spending \CostLF{}. Denoting this value by $A_{\mathrm{alt}}(\DesignVector)$, the relevance of evaluating low fidelity alone is quantified as
\begin{equation}
\begin{aligned}
R_{\LF}(\DesignVector)&=
\begin{cases}
\dfrac{\PositivePart{A_{\mathrm{alt}}(\DesignVector)-A_{\mathrm{H}\setminus\mathrm{L}}(\DesignVector)}}{A_{\mathrm{alt}}(\DesignVector)}, & A_{\mathrm{alt}}(\DesignVector)>0,\\[4pt]
0, & A_{\mathrm{alt}}(\DesignVector)=0,
\end{cases}\\[6pt]
R_{\LF}(\DesignVector)&>\lambda_{\LF}, \qquad
\lambda_{\LF}=\frac{\CostLF}{\CostLFHF}.
\end{aligned}
\label{eq:mfcas_gate}
\end{equation}
Here, $A_{\mathrm{alt}}$ is the maximum among the other affordable opportunities for obtaining a high-fidelity response, either by completing an existing low-fidelity design or through a new nested evaluation. Thus, $R_{\LF}$ measures how strongly the candidate design loses competitiveness relative to the available alternatives when its low-fidelity interpolation uncertainty is removed. The low-fidelity evaluation is separated from the high-fidelity evaluation only when this effect exceeds the fraction of the complete nested-evaluation cost associated with the low-fidelity stage. The criterion therefore introduces no additional acquisition parameter.

Equation~\eqref{eq:mfcas_gate} determines which evaluation is admissible at each new candidate design. When the criterion is satisfied, only the low-fidelity model is evaluated and its actual cost is \CostLF{}; otherwise, the low- and high-fidelity models are evaluated together at cost \CostLFHF{}. At locations already sampled at low fidelity, high-fidelity completion is considered at cost \CostHF{}. MF-CAS then compares all admissible candidate-design and model-evaluation combinations using their corresponding opportunity scores and selects the global maximum. Depending on the information already available and the screening criterion, the selected action may therefore be a high-fidelity evaluation of a design previously evaluated at low fidelity, co-located low- and high-fidelity evaluations of a new candidate design, or a low-fidelity evaluation alone. After each evaluation, the surrogate and uncertainty decomposition are updated and the complete decision is repeated. High-fidelity opportunities therefore provide the reference for advancing the optimization, while low-fidelity-only evaluations are introduced when they provide an intermediate step for reassessing whether a new candidate design remains sufficiently competitive to warrant subsequent high-fidelity evaluation.

\section{Hydrodynamic shape optimization problem}\label{sec:designprob}
\begin{figure}[tbp]
    \centering
    \manuscriptfigure[width=1\textwidth]{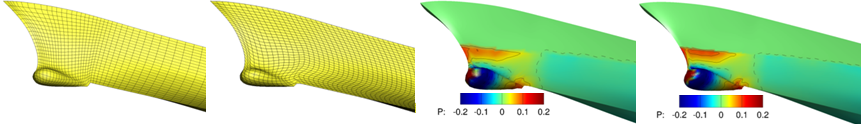}
    \caption{Low- and high-fidelity panel distributions and corresponding pressure fields.}
    \label{fig:lf-hf}
\end{figure}

The frigate DTMB 5415 bare hull form (without appendages) is selected as the design benchmark \cite{serani2026l2sea}. The objective is to minimize the total resistance while satisfying the geometric and hydrostatic requirements imposed on the hull form \cite{pellegrini2022derivativefree}. The optimization problem is formulated as
\begin{equation}
\label{eq:shape_optimization}
\begin{array}{rl}
\mathrm{minimize}
& R_{\rm T}(\DesignVector) \\[1mm]
\mathrm{subject\ to}
& L_{\rm pp}=L_{\rm pp_0},\\
& \nabla(\DesignVector)\geq\nabla_0,\\
& \left|B(\DesignVector)/B_0-1\right|\leq0.05,\\
& \left|T(\DesignVector)/T_0-1\right|\leq0.05,\\
& V_d(\DesignVector)\geq V_0,\\
& \mathbf{x}_l\leq\DesignVector\leq\mathbf{x}_u .
\end{array}
\end{equation}
The design vector $\DesignVector\in\mathbb{R}^{14}$ defines the hull-form modifications, with $\mathbf{x}_l$ and $\mathbf{x}_u$ denoting its lower and upper bounds. The length between perpendiculars $L_{\rm pp}$ is fixed to the reference value, while the displacement $\nabla$ cannot decrease below that of the original hull. Beam $B$ and draft $T$ variations are limited to $\pm5\%$, and the volume $V_d$ reserved for the sonar within the bow dome must remain above its reference value. Subscript "0" denotes the values with the original DTMB 5415 geometry. Hydrodynamic evaluations are performed using WARP, the linear potential-flow solver developed at CNR-INM \cite{WARPCNRINM}. 
The fidelity levels are generated by changing the panel discretization used to represent the hull surface, so that both models describe the same hydrodynamic problem at different spatial resolutions and computational costs. Figure~\ref{fig:lf-hf} compares the low- and high-fidelity panel distributions together with the corresponding pressure fields, illustrating the differences introduced by the two discretizations.

\section{Results and discussion}\label{sec:results}

\begin{figure}[tbp]
    \centering
    \manuscriptfigure[width=0.45\textwidth]{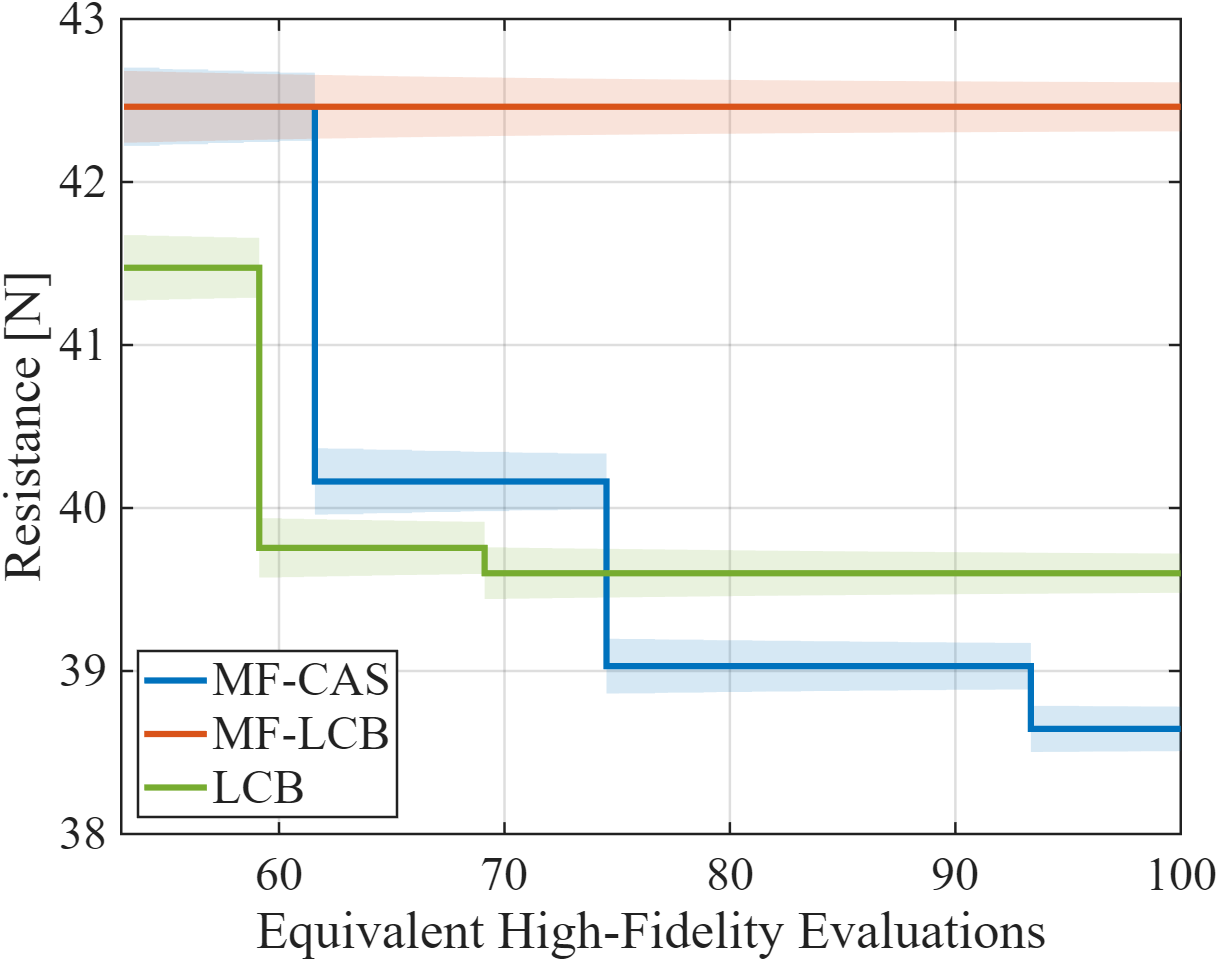}
    \caption{Median best high-fidelity resistance value as a function of equivalent high-fidelity evaluations over five independent runs. Shaded regions indicate the corresponding confidence intervals.}
    \label{fig:convergence}
\end{figure}

The optimization performance is assessed over five independent runs for each method, using the same initial designs and computational budget. Figure~\ref{fig:convergence} reports the median best high-fidelity resistance and the corresponding confidence intervals. Computational expenditure is expressed in equivalent high-fidelity evaluations (eqHF): a high-fidelity evaluation has unit cost $\CostHF=1$, while each low-fidelity evaluation has cost $\CostLF=0.57$ contributes according to its cost relative to the high-fidelity model. The total budget is limited to 100 eqHF, providing a deliberately restrictive setting for the 14-dimensional constrained problem and representative of a preliminary design phase in which only a limited number of expensive simulations can be afforded.

MF-CAS is compared with two baselines using the same uncertainty-aware surrogate formulation. The single-fidelity LCB baseline evaluates only the high-fidelity model. MF-LCB instead uses both fidelity levels: the candidate design is first selected from the high-fidelity LCB, after which the fidelity level is chosen based on the corresponding uncertainty contributions, normalized by their computational costs. The comparison therefore isolates the effect of the active learning policy while retaining the same surrogate and uncertainty formulation. The initial designs are generated using Latin hypercube sampling: the multifidelity initial design of experiment comprises 40 low-fidelity and 30 high-fidelity evaluations, while the single-fidelity search is initialized with 52 high-fidelity evaluations. 

Starting from a common median best resistance of $42.53$ N, MF-CAS reaches $38.6$ N at the end of the available budget, corresponding to a resistance reduction of approximately $9.0\%$. The single-fidelity baseline reaches approximately $39.6$ N, corresponding to a $6.7\%$ reduction, while MF-LCB produces negligible improvement over the initial design. MF-CAS also reduces the resistance below $40$ N after approximately 75 eqHF, compared with approximately 88 eqHF for the single-fidelity baseline, and subsequently reaches values close to $39$ N that are not attained by either baseline within the prescribed budget. This difference is also reflected in the allocation of model evaluations: over the adaptive phase, MF-CAS performs on average 41 low-fidelity and 32 high-fidelity evaluations, whereas MF-LCB allocated the entire budget to low-fidelity, with 80.6 low-fidelity evaluations and no additional high-fidelity evaluations. In MF-CAS, the low-fidelity evaluations comprise 18 low-fidelity only and 23 evaluations performed jointly with high-fidelity, directly illustrating how the proposed policy prevents low-fidelity exploration from dominating the search. 

The different behavior of the two multifidelity policies is consistent with the interaction between the conservative uncertainty estimate and their use of low-fidelity information. The baseline uncertainty promotes exploration in sparsely sampled regions, while MF-LCB subsequently favors the fidelity with the largest uncertainty contribution relative to its cost. Since low-fidelity evaluations are inexpensive, this can lead to extensive low-fidelity exploration before the high-fidelity search progresses. MF-CAS avoids assigning low-fidelity evaluations an independent opportunity proportional to their uncertainty divided by \CostLF{}. Instead, a new candidate design is ranked according to the high-fidelity optimization opportunity and the total cost required to obtain a nested high-fidelity response, while the low-fidelity evaluation may be performed separately as an intermediate step. This prevents the low cost of the LF model from dominating the allocation criterion and provides a more balanced progression between exploration and high-fidelity improvement, consistent with the faster convergence observed in Fig.~\ref{fig:convergence}.

\section{Concluding remarks}\label{sec:remarks}

MF-CAS introduces a multifidelity active learning strategy that uses decomposed surrogate uncertainty to determine when a low-fidelity evaluation is sufficiently informative to precede a more expensive high-fidelity evaluation. Applied to the constrained DTMB 5415 hydrodynamic design problem under a deliberately limited computational budget, MF-CAS achieved a resistance reduction of approximately $9.0\%$, compared with $6.7\%$ for the single-fidelity baseline, while MF-LCB showed negligible improvement. The results indicate that conventional MF-LCB can lose the benefit of multifidelity learning when low-fidelity evaluations are repeatedly selected from their uncertainty-to-cost ratio without considering their influence on the subsequent high-fidelity decision. MF-CAS recovers this benefit by using low-fidelity information selectively to resolve consequential high-fidelity decisions, resulting in a more effective use of the available computational resources.


\begin{thebibliography}{99}
\raggedright

\bibitem{serani2021urans}
Serani, A., Diez, M., van Walree, F. and Stern, F. (2021). URANS analysis of a free-running destroyer sailing in irregular stern-quartering waves at sea state 7. \textit{Ocean Engineering}, 237, 109600.

\bibitem{serani2024scoping}
Serani, A., Scholcz, T. P., and Vanzi, V. (2024). A Scoping Review on Simulation-Based Design Optimization in Marine Engineering: Trends, Best Practices, and Gaps. \textit{Archives of Computational Methods in Engineering}, 31(8), 4709-4737.

\bibitem{mainini2024lecture} 
Mainini, L. Advanced Multifidelity Active Learning: Formulations and Applications to Engineering Design Problems in Aeronautics and Space. AVT-385 RLS on Multi-Fidelity Methods for Multidisciplinary Design Optimization. STO Research Lecture Series. 1 July 2024. 

\bibitem{mainini2026arco} 
Mainini, L., Serani, A., Pehlivan Solak, H., Di Fiore, F., Rumpfkeil, M. P., Minisci, E., Quagliarella, D., Yildiz, S., Ficini, S., Pellegrini, R., Thelen, A., Bryson, D., Nikbay, M., Diez M. and Beran, P. S. (2026). Analytical benchmark problems and methodological framework for the assessment and comparison of multifidelity optimization methods. \textit{Archives of Computational Methods in Engineering}, 33(2), 2969-3000.

\bibitem{volpi2015development} 
Volpi, S., Diez, M., Gaul, N. J., Song, H., Iemma, U., Choi, K. K., Campana, E. F. and Stern, F. (2015). Development and validation of a dynamic metamodel based on stochastic radial basis functions and uncertainty quantification. \textit{Structural and Multidisciplinary Design Optimization}, 51(2), 347-368.

\bibitem{pellegrini2023multifidelity} 
Pellegrini, R., Wackers, J., Broglia, R., Serani, A., Visonneau, M. and Diez, M. (2023). A multi-fidelity active learning method for global design optimization problems with noisy evaluations. \textit{Engineering with Computers}, 39(5), 3183-3206.

\bibitem{wackers2024adaptivity} 
Wackers, J., Pehlivan Solak, H., Pellegrini, R., Serani, A. and Diez, M. (2024). Adaptivity and uncertainty of multi-fidelity surrogate models for shape optimization. In \textit{16th World Congress in Computational Mechanics}.

\bibitem{karen2018siam}
Peherstorfer, B., Willcox, K. and Gunzburger, M. (2018). Survey of multifidelity methods in uncertainty propagation, inference, and optimization. \textit{SIAM Review}, 60(3), 550-591.

\bibitem{pellegrini2022derivativefree} 
Pellegrini, R., Serani, A., Liuzzi, G., Rinaldi, F., Lucidi, S. and Diez, M. (2022). A derivative-free line-search algorithm for simulation-driven design optimization using multi-fidelity computations. \textit{Mathematics}, 10(3), 481.

\bibitem{difiore2023nonmyopic} 
Di Fiore, F. and Mainini, L. (2023). Non-myopic multipoint multifidelity Bayesian framework for multidisciplinary design. \textit{Scientific Reports}, 13(1), 22531.

\bibitem{difiore2024nm2bo} 
Di Fiore, F. and Mainini, L. (2024). NM2-BO: Non-Myopic Multifidelity Bayesian Optimization. \textit{Knowledge-Based Systems}, 299, 111959.

\bibitem{takeno2020multifidelity} 
Takeno, S., Fukuoka, H., Tsukada, Y., Koyama, T., Shiga, M., Takeuchi, I. and Karasuyama, M. (2020). Multi-fidelity Bayesian optimization with max-value entropy search and its parallelization. \textit{International Conference on Machine Learning}

\bibitem{ruan2020variablefidelity} 
Ruan, X., Jiang, P., Zhou, Q., Hu, J. and Shu, L. (2020). Variable-fidelity probability of improvement method for efficient global optimization of expensive black-box problems. \textit{Structural and Multidisciplinary Design Optimization}, 62(6), 3021-3052.

\bibitem{WARPCNRINM} 
Bassanini, P. (1994). The wave resistance problem in a boundary integral formulation. Surv. Math. Ind, 4, 151-194.

\bibitem{serani2022hullform} 
Serani, A., Stern, F., Campana, E. F., and Diez, M. (2022). Hull-form stochastic optimization via computational-cost reduction methods. \textit{Engineering with Computers}, 38(3), 2245-2269.

\bibitem{serani2026l2sea} 
Serani, A. and Diez, M. (2026, March). L2 Sea Benchmark. https://doi.org.10.5281/zenodo.19351758
\end{thebibliography}
\end{document}